\documentclass[12pt,preprint]{aastex}
\def\alwaysmath#1{\ifmmode{#1}\else{$#1$}\fi}
\def\msun{\alwaysmath{{M}_{\odot}}}

\def\arcsec{\hbox{$^{\prime\prime}$}}
\def\etal{{et al.~}}

\newcommand\HST{{\sl HST}}
\newcommand\hst{{\sl HST}}

\def\ltsima{$\; \buildrel < \over \sim \;$}
\def\gtsima{$\; \buildrel > \over \sim \;$}
\def\lsim{\lower.5ex\hbox{\ltsima}}
\def\gsim{\lower.5ex\hbox{\gtsima}}

\slugcomment{In press in ApJ}

\begin{document}

\title{Discovery of another peculiar radial distribution 
of Blue Stragglers in Globular Clusters: The case of 47 Tuc\altaffilmark{1}} 

\author{Francesco R. Ferraro\altaffilmark{2}, 
Giacomo Beccari\altaffilmark{2}, Robert T. Rood\altaffilmark{3}, 
Michele Bellazzini\altaffilmark{4},
 Alison Sills\altaffilmark{5},
Elena Sabbi\altaffilmark{2}
}

\altaffiltext{1}{Based on
observations with the NASA/ESA HST, obtained at
the Space Telescope Science Institute, which is operated by AURA, Inc.,
under NASA contract NAS5-26555. Also based on WFI
observations collected at the European Southern Observatory,
La Silla, Chile, within the observing
programs 62.L-0354 and 64.L-0439.}
\altaffiltext{2}{Dipartimento di Astronomia Universit\`a 
di Bologna, via Ranzani 1, I--40127 Bologna, Italy,
ferraro@bo.astro.it}
\altaffiltext{3}{Astronomy Dept., University of Virginia, Charlottesville,  
VA 22903-0818, USA; rtr@virginia.edu }
\altaffiltext{4}{INAF -  Osservatorio Astronomico
 di Bologna, via Ranzani 1, I--40127 Bologna, Italy,
 bellazzini@bo.astro.it}
\altaffiltext{5}{Department of Physics and Astronomy, McMaster University, 
 1280 Main Street West, Hamilton, ON, L8S 4M1, Canada;
 asills@mcmaster.ca}
 
\begin{abstract}

We have used high resolution {\it WFPC2-HST} and wide field
ground-based observations to construct a catalog of blue straggler
stars (BSS) in the globular cluster 47 Tuc spanning the entire radial
extent of the cluster.
   
The BSS distribution is highly peaked in the cluster center, rapidly
decreases at intermediate radii, and finally rises again at larger
radii. The observed distribution closely resembles that discovered in
M3 by Ferraro et al (1993,1997).  To date, complete BSS surveys
covering the full radial extent (\hst\ in the center and wide field
CCD ground based observations of the exterior) have been performed for
only these two clusters.  Both show a bimodal
radial distribution, despite their different dynamical
properties.  BSS surveys covering the full spatial extent of
more globular clusters are clearly required to determine how common
bimodality is and what its consequence is for theories of BSS
formation and cluster dynamics.
\end{abstract}
 
 
\section{Introduction} 
\label{sec:intro}
Blue straggler stars (BSS) are found primarily in star clusters, where
they appear as a sparsely populated extension of the main sequence
above the turnoff point. Superficially they appear to be main sequence
stars with masses larger than expected for the cluster at its turnoff
age. There are two proposed mechanisms thought to produce
BSS: the first is mass exchange in a binary system and the second is
the merger of two stars induced by stellar interactions 
(either single or in binaries) in a
dense stellar environment. Globular cluster cores are obvious targets
for BSS not only because of their high stellar density, but because
primordial binaries in the clusters may well have sunk to the cores.

With the advent of the Hubble Space Telescope (\hst) it became
possible for the first time to search dense cluster cores for
BSS. Searches in the ultraviolet (UV) can be particularly effective,
not because BSS are especially bright in the UV, but because the red
giants are faint in the UV. The photometric blends which mimic BSS in
visible CMDs are far less problematic in the best UV CMD planes, and
it is possible to obtain complete BSS samples in even the densest
cores. With \hst, the densest cluster cores are {\it the obvious
place} to search for BSS, and there have been few systematic searches
for BSS in the outer parts of clusters in recent years. Indeed only
one cluster, M3, has been adequately surveyed for BSS over its entire
radial extent \citep{f93,f97}.

Since it is a massive cluster with high central density, 47~Tuc
is an
obvious target for a BSS search, and its core has been the subject of
several earlier investigations
\citep{gua92,demarchi47bss,ferraro47bss,knigge47}. These studies have
covered only a fraction of the core either because they used the Faint
Object Camera or just the PC of WFPC2. Albrow et al. (2001) found 
BSS in their WFPC2 search for eclipsing biniaries in the core of
47~Tuc. Blue stragglers have also been
studied outside the core of 47 Tucanae by Sills et al. (2000) who also
 modeled the BSS formation rate. Kaluzny et al. (1998)
note several BSS candidates in their CMD of 47~Tuc, and there are
clear BSS candidates in the Str\"omgren photometry of
Grundahl et al. (2002). Only a small part of the cluster exterior was
covered by these studies.

With new wide field imagers on ground based telescopes at sites with
excellent seeing it is now possible to perform BSS surveys 
which yield
complete samples over the entire extent of the cluster. 
 Here we present results covering both the full cluster core using
\hst\ WFPC2, and the exterior of the cluster using the wide field
imager at the ESO 2.2\,m telescope.

\section{Observations and  data analysis}
\label{sec:obs}

\subsection{The data-set}

The photometric data used here consist of two sets: 

{\it (i)---High
resolution set}---a series of public HST/WFPC2 images taken through
the UV filter F218W and F439W ($B$).  The images have been retrieved
from the {\it ESO/ST-ECF Science Archive}, the automatic {\it Archive
Association Procedure} which allows retrieval of a set of images taken
in the same filter and in similar conditions. The images are
automatically combined and cleaned of cosmic rays.  The first part of
this data-set (specifically the PC) has been already published in
Ferraro et al (2001). Here we present the analysis of the entire
data-set (PC, WF2, WF3, WF4). The information about the observing
epochs, the proposal ID number, the filter, and the number of
exposures can be found in Table 1 of Ferraro et al 2001.  The
photometric reductions of the high resolution images were carried out
using ROMAFOT (Buonanno et al. 1983), a package developed to perform
accurate photometry in crowded fields and specifically optimized to
handle under-sampled point spread function (PSF) (Buonanno \&
Iannicola 1989) as in the case of the HST-WF chips.

PSF-fitting instrumental magnitudes have been obtained using the
standard procedure described in Ferraro et al. (1997a, 2001).  The
final catalog of the F218W and F439W magnitudes was calibrated by
using the zero-points listed by Holtzman et al. (1995).
   
{\it (ii)---Wide Field set---} a complementary set of  multi-filter
($B$, $V$, $I$) wide field images was secured during an observing run at
the 2.2m ESO-MPI telescope at ESO (La Silla) in July 1999, using the
Wide Field Imager (WFI).  An additional data-set of public WFI images
  were also
retrieved from the ESO/ST-ECF Archive. The WFI has exceptional imaging
capabilities---each image consists of a mosaic of 8 CCD chips (each
with a field of view of $8'\times 16'$) giving a global field of view
of $33'\times 34'$.
 
The raw WFI images were corrected for bias and flat field, and the
over-scan region was trimmed using standard IRAF\footnote{IRAF is
distributed by the National Optical Astronomy Observatory, which is
operated by the Association of Universities for Research in Astronomy,
Inc., under cooperative agreement with the National Science
Foundation.} tools. 
 The PSF fitting procedure
was performed independently on each $V$, $B$ and $I$ image, using
DoPhot \citep{dophot}.    For each passband at least 7
exposures were combined, and the resulting magnitudes transformed to
the same (instrumental) photometric system; stars
in common were then averaged with the requirement that
only stars successfully measured at least in 3 images
per filter were included in the final catalog. 
 A final catalog listing the instrumental $B,~V,~I$
magnitudes for all the stars in each field was finally
obtained.

\subsection{Astrometry}

The entire catalog HST+WFI has been placed on the absolute astrometric
system by adopting the procedure already described in other papers
(see Ferraro et al 2001, 2003).  The new astrometric {\it Guide Star
Catalog} ($GSCII$) recently released and now available from the web
({\rm http://www-gsss.stsci.edu/gsc/gsc2/GSC2home.htm}),
was used to search for astrometric standard stars lying in the WFI
image field of view. Several hundred $GSCII$ reference stars have been
found in the WFI FoV, allowing an accurate absolute positioning of the
image.  In order to derive an astrometric solution for each of the 8
WFI chips, we used a program specifically developed at the Bologna
Observatory (P. Montegriffo et al 2003, in preparation). 
 Then, a few hundred stars in the overlapping area
between the WFI and the WFPC2 field of view were used as secondary
astrometric standards in order to place the HST catalog in the
absolute astrometric system. At the end of the entire procedure the
rms residuals were of the order of $\sim 0\farcs 3$ both in RA and Dec
and we take this value as representative of the astrometric
accuracy.

We used our standard technique (Montegriffo et al 1995, Ferraro et al
1999, 2001, 2003) to find the center of gravity ($C_{\rm grav}$).  The
barycenter of resolved stars is compatible (within 1--2 arcsec) with
the center of luminosity $C_{\rm lum}$ listed by Djorgovski \& Meylan
(1993). Thus in the following we use their value as center:
$\alpha_{\rm J2000} = 00^{\rm h}\, 24^{\rm m}\, 05\fs20,
~\delta_{J2000} = -72\arcdeg\, 04\arcmin\, 51\farcs$.

\section{Definition of the samples}  
 
\begin{figure} 
\plotone{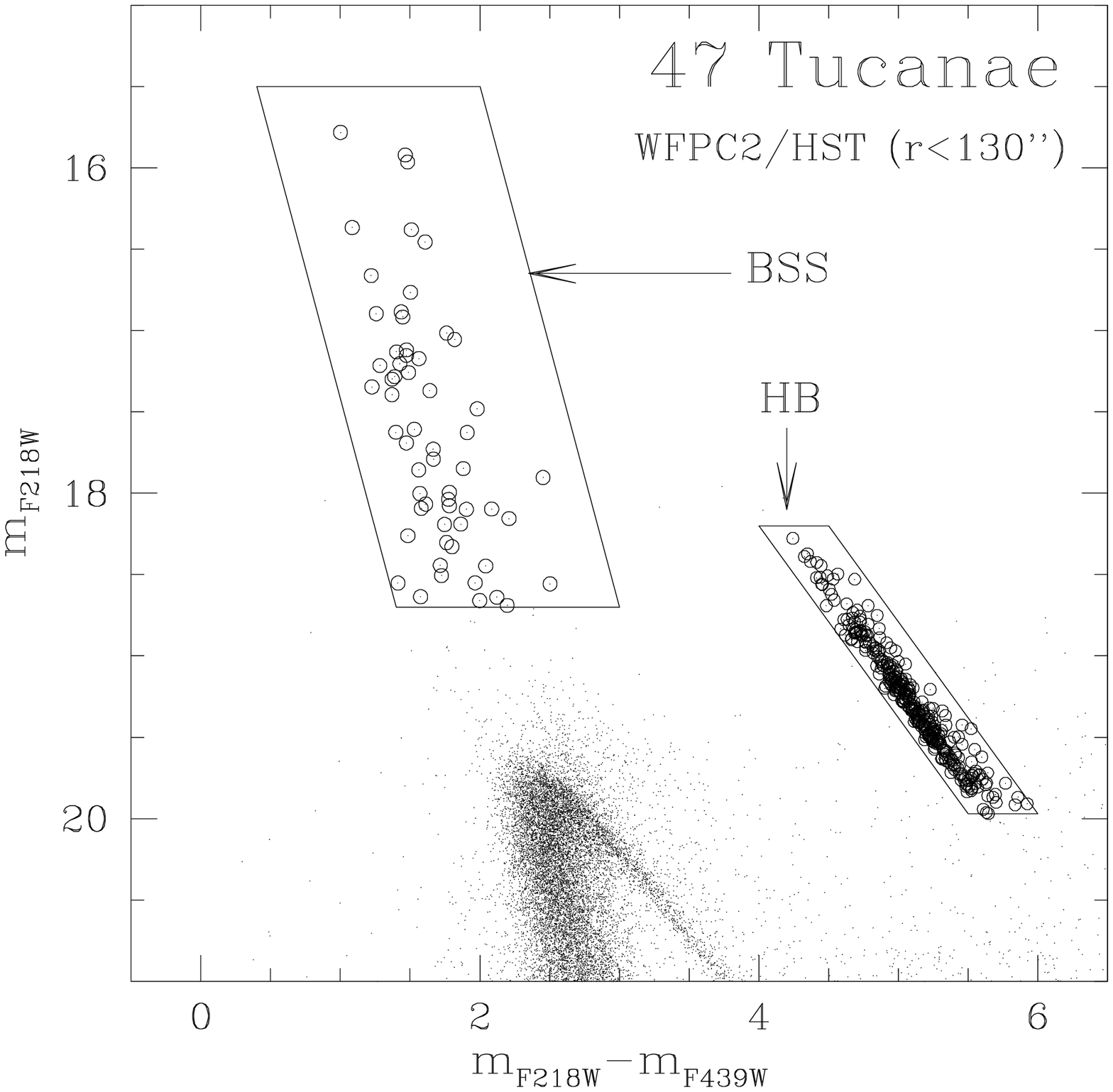} 
\caption[fig1.ps]{
 ($m_{218}, m_{218}-m_{439}$) CMD for the 
central region of 47 Tuc ($r<130''$) from
WFPC2/HST observations.
  The two selection boxes for the BSS and the
 reference (HB) population are shown.
\label{fig:CMD1}}
\end{figure}

\begin{figure} 
\plotone{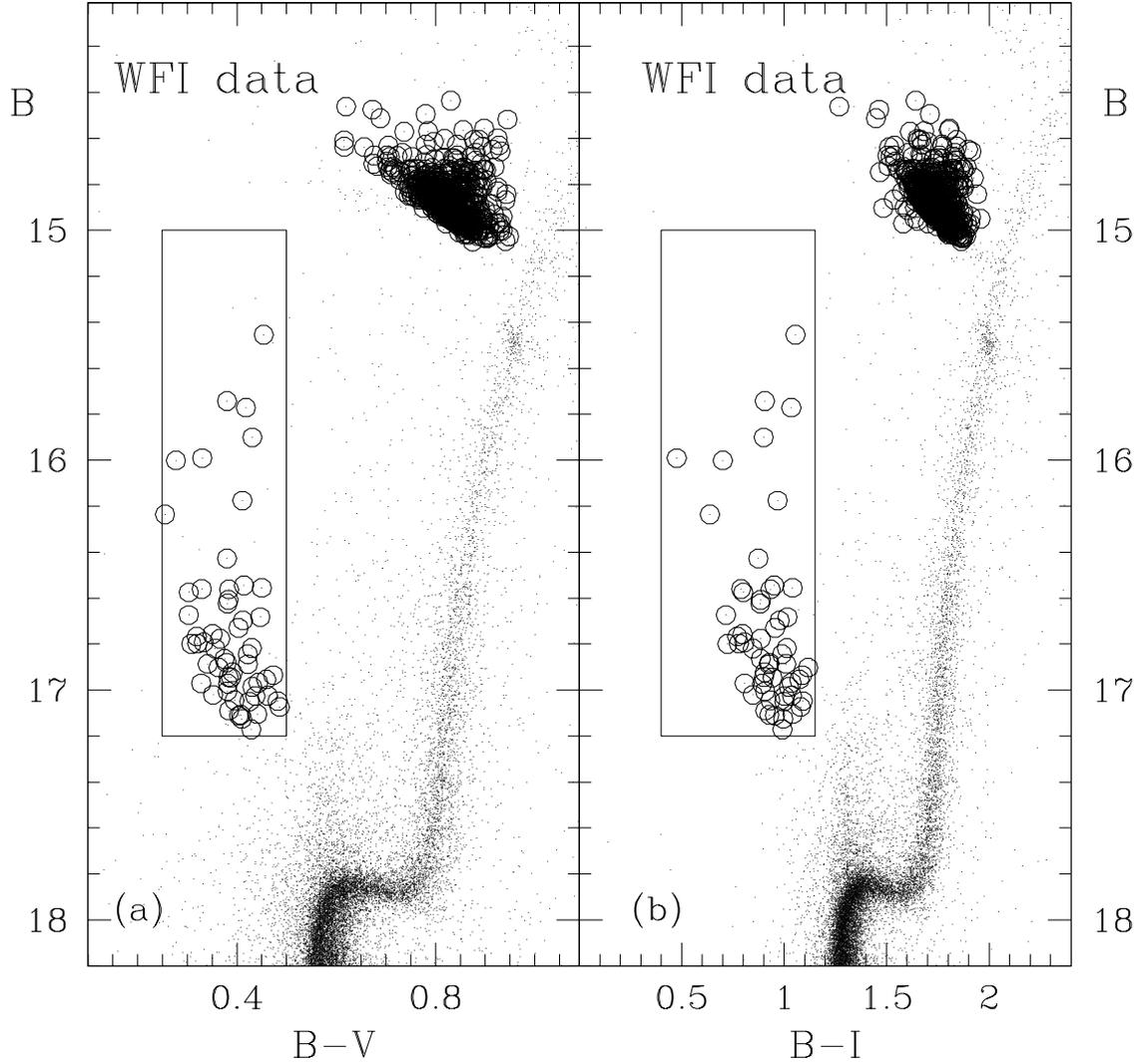} 
\caption[fig1.ps]{$B,B-V$ ({\it Panel (a)})
and $B,B-I$ ({\it Panel (b)})
CMDs for the external part
 of 47 Tuc ($130''<r<1500''$) from ground-based
 (WFI) observations. As in Figure 1 stars selected
 in the  BSS and in the  reference (HB) 
 population are marked
   with empty circles.
\label{fig:CMD2}}
\end{figure}

In order to avoid spurious effects due to incompleteness of the ground
based observations in the crowded central region of the cluster we
restricted the WFI sample to the outer region $r>130\arcsec$ (hereafter
the {\sl WFI sample}). In the inner $r<130\arcsec$ our sample includes
only stars in the WFPC2 FoV (hereafter the {\sl HST sample}). This
sample takes advantage of the superior capability of UV observations
in detecting BSS (Ferraro et al 1999, 2001) in the
overcrowded central regions of high density clusters. The Color-Magnitude
Diagrams (CMDs) derived from these samples are shown in
Figure~\ref{fig:CMD1} and Figure~\ref{fig:CMD2}.

\subsection{ The BSS selection}

Since the two data-set have been observed in different
 photometric bands, different selection boxes have to be
 defined to isolate the populations in the two CMDs.
The  definition of the BSS sample is
 quite natural  in the UV-CMD (Figure 1) since,
  as usually seen, in these planes 
  the BSS define a quite distinct sequence spanning $\sim 3$
magnitudes.  A total of $\sim 100$ BSS have been identified in the
HST-WFPC2 FoV. However, as in our previous studies of M3 and M80
(respectively, Ferraro et al 1997, hereafter, F97, and Ferraro et
al 1999, hereafter, F99) to avoid any possible contamination due to
blends, incompleteness etc, in the following we limit our analysis to
the brightest portion of the BSS sequence. 
The faint threshold of the BSS selection box has been chosen to be
$m_{218}<18.7$. This limit is nearly 1 magnitude
brighter than  the
cluster Main Sequence (MS) Turn-Off (TO)
 in the UV plane to avoid any
spurious effect due to blends.
The adopted selection box
for the BSS in the HST sample is shown in Figure 1.  59
Bright-BSS (b-BSS) have been counted in this selection box. 
 
In order to make a meaningful BSS selection in the WFI sample we
converted the magnitude extention of the box
defined in the UV plane into the
ground-based ($B,~B-V$) and ($B,~B-I$) planes. The
corresponding values have been computed by using the $B$ filter which is
in common in the two data sets (since the F439W WFPC2-filter is
approximately a $B$ filter). In doing this we first aligned the two CMDs
in the ($m_{F439W},m_{F218W}-m_{F439W}$) and ($B,~B-V$) planes, by using
the the SGB and the HB level as reference, and then we determined the
distribution of the selected bright BSS in $B$ magnitude.  This
procedure gives a value of $B\sim 17.3$ and $B\sim 15$  
for the faint and the bright threshold (respectively) of the
BSS box in the WFI sample.

The color limits of the BSS box in the ground-based planes have been
chosen to minimize the possibility of field contamination. 
The red edge has been assumed at $B-V=0.5$, where a significative
decrease of the galaxy field contribution is expected (see for example
Figure 2 by Unavane et al (1996) and Figure 4 in Lemon et al.,
2003). The blue edge has been choosen in order to exclude a few blue
objects which lie significatly outside the BSS sequence (see Figure 2
(a)).  This selection includes the bulk of the BSS population, while a
few evolved BSS might be lost in the region dominated by
the field population. However, as discussed in previous papers, the
number of lost BSS is certainly small (see as example the case of M3
(Ferraro et al 1993, 1997)). We emphasize that even slightly
different assumptions in the definition of the selection boxes would
not alter the main results presented in this paper. With these limits
an appropriate selection box for the bright BSS has been defined in
the both the optical planes (see Figure 2- panel (a) and (b)).  A
preliminary selection of possible BSS has been done in the ($B,~B-V$)
plane, but the the final selection has been done by using both
($B,~B-V$) and ($B,~B-I$) CMDs. Thus only the 57 objects which lie in
the BSS selection boxes in both the CMDs have been selected.

Figure 3 shows the final comparison of the BSS samples in the two
data-set (HST and WFI, respectively) considered below. In order to
make the two samples completely homogeneous, the faintest 6 BSS in the
HST sample which fall just below the adopted threshold ($B\sim 17.3$)
have been excluded. The final sample consists of 53 BSS in the
HST sample and 57 BSS in the WFI sample (for a total of 110
objects).

 \subsection{ The reference population} 
 
 In order to study the behavior of the
BSS with respect to the normal stars in the cluster, we have to select
a reference population which is assumed to trace the cluster stars. As
discussed in previous papers (see for example Ferraro et al 2003) the
most {\it natural} reference population in the UV-CMDs is the
Horizontal Branch (HB), since it is well separated from the other
branches and HB stars are quite bright in most CMD planes (see Figure
1).

The selection of the HB is quite easy in both planes since
it defines a well defined visible clump of stars in both
the UV (see Figure 1 and also Figure 2 in Ferraro et al
2001) and the ground-based CMDs (Figure 2).  Again the HB
selection is defined independently in all the CMDs and is
designed to include the bulk of the HB population. The HB
box shown in Figure 1 and the selection shown in Figure 2
assure the inclusion of the bulk of the HB stars: small
differences in the assumptions on the shape of the box
would include or exclude a few stars. For example, a
fainter level for the HB box in the UV plane would include
a few scattered stars which might be RGB stars. In the
ground-based planes (Figure 2) we designed an appropriate
box to avoid the inclusion a small sequence of
underluminous HB stars which are clearly lie below the ZAHB
in Figure 2(a). Again the inclusion of these stars (which
are less than a few percent of the global HB population)
does not affect our conclusions in any way.

   Following the criteria described above a total sample
   of 314 and 581 HB stars have been counted in the HB
selection boxes, in the HST and WFI sample, respectively.

 The global sample of 116 bright BSS and 895 HB stars spans almost
the entire radial extension of the cluster ($0$ up to $23'$ from the
cluster center) and is the largest sample ever published in 47~Tuc.

\subsection{Field Contamination}
 
  Galaxy foreground contamination in the BSS region is
  negligible thanks to the selection in color ($B-V<0.5$)
  see Figure 4 in Lemon et al (2003) and figure 2 by
  Unavane et al (1996).  By using the star counts by
  Ratnatunga \& Bahcall (1985)  we estimate that $\sim 100$
  field stars with  $B-V<0.8$ and $13<V<17$ are expected in
  the WFI FoV. Figure 2 by  Unavane et al (1996) and Figure
  4  by  Lemon et al (2003) showed that the bulk of the
  Galaxy foreground  stars are expected to have $B-V>0.5$.

  The brigher portion of the RGB  of the Small Magellanic
  Cloud (SMC) field population  is clearly visible in our
  CMD at $B=18.2,~ B-V=1.6$. Moreover the main contribution
  of the SMC background stars is expected  at magnitudes
  significantly fainter than the BSS selection box (see
  Figure 1 by Zoccali et al. 2001). In any case the BSS
  selection box is located in the region between the  Main
  Sequence (MS) of young stars and the evolved RGB stars
  (compare the BSS selection box and Figure 10 by Momany et
  al, 2001)  where only a few SMC field stars  are
  expected. In accord with the above considerations no
  correction has been applied to the BSS counts.

\section{Results}
 
As a first step the radial distribution of the 110 b-BSS has been
compared to that of the reference population. The cumulative radial
distribution for the 110 b-BSS and the 895 HB stars are plotted in
Figure 4 (panel (a)) as a function of the projected distance from the
cluster center. As can be seen the behavior of the two distributions
is not monotonic since the BSS appear to be more concentrated than the
HB in the central region and less concentrated than the HB in the
outer region.
Note that the shoulder visible in
the radial distribution shown in Figure 4 (panel (a))
is due to the HST/Ground-based
sample transition. As discussed above we conservately 
decided to consider in the region $r<130
\arcsec$ only stars observed with HST. Because of
the shape of the WFPC2 FoV  there is
 a region at $r\sim 110-130$ where only 
a few stars can be detected. This region is indicated by a
vertical shaded region in Figure 4 (panel (a)). The
assuption of using only the HST sample
in the most inner region of the cluster
 is, in our opinion,  the safest possible choice, since 
 the ground based magnitudes  could be still severely affected by
 the crowding conditions at this distance from the cluster
 center. Note that this shoulder effect vanishes when the BSS
 population is scaled to a reference population (HB or RGB) or
 fractional luminosity as we do below.
 
The effect is better illustrated in Figure 4 (panel (b)) in which the
radial distribution for two sub-samples HST and WFI are shown
separately. As can be seen the BSS are clearly more concentrated in
the central region (HST sample) and less concentrated in the outer
region (WFI sample) with respect to the reference HB population.  A
Kolmogorov-Smirnov test has been applied to the distributions to check
the statistical significance of the difference. The test yields a
probability of $99.96\%$ (more than $ 3 \sigma$ level of confidence)
and $98.86\%$ ( $\sim 2.5 \sigma$ level of confidence) that the b-BSS
population has a different radial distribution with respect to the HB
stars.  This distribution closely resembles that already observed in
M3 (F93, F97): indeed Figure 4 (panel (b)) is an almost exact copy of Figure
7 in F97.
We accordingly follow the procedure described
in F93 (applied also in F97) where the surveyed area has been divided
into a set of concentric annuli.  Thus, 11 concentric annuli (each
containing roughly $\sim 10\%$ of the reference population) have been
defined (see Table 1).  The number of BSS and HB stars have been then
counted in each annulus.  The first observable quantity that
can be measured in each annulus is the ratio $N_{\rm BSS}/N_{\rm HB}$.
 
Figure 5 (panel (a)) shows the distribution of this ratio as a
function of the effective radius of each annulus. The distribution is
clearly bimodal with the highest value in the innermost annulus. From
this quite large value ($N_{\rm BSS}/N_{\rm HB}\sim 0.4$) the ratio
significantly decreases to less than 0.1 as $r$ increases and then
slowly rises up in the outer region up to $\sim 0.3$.  Again this
trend closely resembles that discovered in M3 by F97. In order to
better analyze this trend we have computed the {\it relative
frequency} of BSS and HB normalized to the fraction of luminosity
sampled in each annulus. 
The fraction of light sampled  in each
annulus has been computed from the cluster
brightness profile, and appropriately scaled to the  the
fraction of area covered by the observations in each
annulus.

The original
definitions by F93 are

\begin{displaymath}
R_{\rm BSS} = {{(N_{\rm BSS}/N_{\rm BSS}^{\rm tot})} \over 
{(L^{\rm sample}/L_{\rm tot}^{\rm sample})}} 
\end{displaymath}

\begin{displaymath}
R_{\rm HB} = {{(N_{\rm HB}/N_{\rm HB}^{\rm tot})} \over 
{(L^{\rm sample}/L_{\rm tot}^{\rm sample})}} 
\end{displaymath}

\noindent

The numbers of BSS and HB in each annulus, the $N_{\rm
BSS}/N_{\rm HB}$ ratio, the sampled luminosity and the
specific frequencies defined above are listed in Table 1.
The {\it relative frequency} of BSS as a function of the
distance from the cluster center is then plotted in Figure
5 (panel (b)) and compared with the corresponding one for
the HB ``reference'' stars. As can be seen, the HB specific
frequency remain essentially constant over the surveyed
area since the fraction of HB stars (as any post-main
sequence stage) in each annulus strictly depends on the
fraction of luminosity sampled in that annulus (see the
relation by Renzini\& Buzzoni  1986, eq 2 in Ferraro et al
2003).  In contrast, the BSS specific frequency reaches its
maximum at the center of the cluster then decreases to an
approximately constant value in the range
100\arcsec--500\arcsec\ from the cluster center and then
rises again. This behavior fully confirms the trend shown
in Figure 5 (panel (a)) and suggests that dynamical 
events and/or formation
mechanism shape the radial distribution of the BSS in the
cluster (M. Mapelli et al. 2003, in preparation) 

To further demonstrate the similarity with M3 and to study possible
differences we show BSS specific frequencies for the two clusters on
the same figure (see Figure 6). The radial coordinate has been
expressed as a function of the core radius $r_c$. Using our observed
star counts we have found values of $r_c=21\arcsec$ and
$r_c=24\arcsec$ for 47 Tuc and M3, respectively.
In order to derive the core radius we followed the
same procedure shown in Ferraro et al (2003b).
 All stars brighter than the cluster MS-TO   have been used
 to derive the star density profile that was then fit by
 the projected star density obtained from a standard
 isotropic single-mass King-model. The full discussion of 
 the star density profile for both (and a few other)
 clusters will be presented in a forthcoming paper (Ferraro
 et al 2003, in preparation). A typical uncertanty of $\pm
 2''$ can be assumed in the $r_c$ determination.
Note  that the value assumed for 47 Tuc is  consistent with
that ($r_c=24\arcsec \pm 2\arcsec$) recently obtained by
Howell, Guhathakurta \& Gilliland (2000) in a  review 
discussing the most recent determinations of the core
radius of 47 Tuc by using HST.  Some points about Figure 6
are worth noting: (1) the central values are similar; (2)
while the BSS specific frequency decreases in both clusters
as $r$ increases from 0 to $\sim 4r_c$ the decrease is much
larger in M3. In 47~Tuc it is a factor of 5.5 dropping from
2.64 down to 0.48; in M3 the drop is a factor of 15 (from
2.76 to 0.2). (3) the main difference in the two
distributions is that the specific frequency minimum in
47~Tuc appears to be much broader than that observed in M3.
In 47 Tuc the depletion zone is extends from $\sim 4 r_c$
to 20--$22 r_c$ and with the upturn of the BSS density
occurring at $\sim 25 r_c$, while in M3 the BSS specific
frequency is already rising at $\sim 8 r_c$.

\section{Discussion}
  
F97 argued that the bimodal distribution of BSS in M3 was evidence
that two formation scenarios were at action in the same cluster: the
Exterior BSS (EBSS) arising from mass transfer in primordial binaries
and the Central BSS (CBSS) arising from stellar interactions which
lead to mergers. As had been earlier noted by Bailyn \& Pinsonneault
(1995), the EBSS and CBSS luminosity functions differed in the sense
theoretically expected for two mechanisms.

One difficulty with primordial binaries as a source for the
EBSS is that one ordinarily would expect that they would
have settled to the cluster centers since relaxation times
are typically less than 1~Gyr \citep{djor-physparm}.  
Simulations to address the question of how many primordial
binaries can form in the external regions of the cluster
and how long primordial binaries with mass approaching
twice the turnoff mass could remain in the outer parts of a
cluster are now in progress (M. Mapelli et al 2003, in
preparation).

Sigurdsson \etal\ (1994) offered another explanation for the bimodal
BSS distribution in M3. They suggested that the EBSS were formed in the
core and then ejected into the outer regions by the recoil from the
interactions. Those binaries which get kicked out to a few $r_c$
rapidly drift back to the center of the cluster due to mass
segregation, leading to a concentration of BSS near the center and a
paucity of BSS in the outer parts of this region.  More energetic
kicks will take the BSS to larger distances; these stars require much
more time to drift back toward the core and may account for the
overabundance of BSS at large distance.

How does the discovery of a bimodal radial distribution of BSS in
47~Tuc affect these arguments? Dynamically 47~Tuc is quite  different
from  M3: its central density is 40 times larger \citep{pm93},
and its core is rich in X-ray binaries, millisecond pulsars, and other
interaction products \citep{ferraro47bss,grindlay01, camilo00,e03}. We
have already shown \citep{bss6} that CBSS populations cannot be
characterized by a simple parameter like collision rate, so the fact
that CBSS in 47~Tuc and M3 have comparable specific frequencies should
not come as a surprise.

It should be noted that an BSS specific frequency upturn similar to
that observed in 47 Tuc and M3 has been detected also in M55 (Zaggia
et al. 1997).  This is of particular significance because M55 has a low
central density---0.1 that of M3. Note that although the data set
presented by Zaggia et al. for M55 are ground based
observations and covered only a quadrant of the cluster, there is at
least preliminary evidence that a upturn is present also in a low
density cluster like M55. 
We note that the effect found by Zaggia et al in M55 could be
even stronger since ground-based observations tend to hide
BSS in the central region of the cluster. For example
 ground based observations of the center of M3
led Bolte et al. (1993) to claim depletion of BSS in the center
whereas \hst\ observations eventually found a peak. M55 is surely less
concentrated than M3, but still intermediate crowding conditions and
the presence of highly saturated giants in the center can produce some
level of incompleteness at the BSS level. 

The implications of the bimodality and relatively large
populations of EBSS are not yet clear beyond the fact that EBSS may be
fairly common and form in diverse clusters. If the BSS bimodality in
M55 is confirmed, we will be faced with the result that clusters with
central densities ranging over a factor of 400 can produce a bimodal
BSS distributions. Any dynamical model like that of Sigurdsson et al. would
seem more applicable to 47~Tuc than M3. It would be a real stretch to
reach the dynamical conditions in M55.

Given these factors it seems even more likely that primordial binaries
play a fundamental role in the BSS formation mechanism. Better
information on the abundance of primordial binaries in clusters and
their survival rate at a given distance from the cluster center is
essential.  In 47 Tuc a recent extensive search for binary stars in
the core (Albrow et al 2001) has led to an overall binary frequency of
$14\%\pm 4\%$. Their Fig.~21 shows that the binaries appear to be
significantly more centrally concentrated with respect to the normal
cluster stars but less concentrated than the BSS. Hence, if the
cluster is in dynamical equilibrium, the CBSS are among the heaviest
visible stellar components in the core of 47 Tuc. Albrow et al. also
note that the W~UMa stars form a sequence in the CMD which like that
which would be expected for equal mass binaries. The brightest lie in
the BSS region of the CMD. Since W~UMa systems are contact binaries
they may well merge to become BSS.

Unfortunately, binary populations have been measured in only a few
additional clusters (most notably NGC~6752, NGC~288), and they
probably have diverse origins: {\it (i)} in the core of NGC~6752
\citep{ruba1} the binaries are probably not primordial, since NGC~6752
is such a dynamically evolved cluster \citep[and references
there-in]{ferraro6752bh}; {\it (ii)} in the low density cluster
NGC~288 \citep{bellazzini288bin}, the binaries are probably
primordial. Despite the much longer relaxation time of NGC~288, the
binaries have all settled to within the inner half light radius; The
technique used by Rubenstein \& Bailyn (1997) and Bellazzini et
al. (2002) requires only modest \HST\ observation time and should be
applicable to any cluster. Even though such studies of the binary
populations are painful, we need more.

Clearly, it is desirable to have full 
BSS surveys for more
clusters. It may well be a fluke that the first two clusters to be fully
surveyed have bimodal distributions. We have pointed out \citep{bss6}
that all properly studied clusters have CBSS. The existence of EBSS in
M3 \citep{sandage53} marked the discovery of BSS. Similarly there is
clear evidence for EBSS in ground based CMDs of 47~Tuc
\citep{sills00,grundahl47,kaluzny47}. Quite possibly all clusters with
significant EBSS populations are bimodal. On the other hand, high
quality CMDs of other clusters like M5 \citep{sandquistm5} and M80
\citep{brocatom80} show evidence for few if any EBSS. If more careful
studies of such clusters do not reveal a yet undetected EBSS
population, we will have to search for a mechanism which produces
bimodal radial BSS distributions in some clusters but not in others.

The population of BSS discovered here opens a new window for the
investigation of the origin and the formation mechanism of BSS in
GGCs.
In fact, this discovery  suggests that the {\it
peculiar} radial distribution firstly found in M3 is much
more {\it common} that was thought. Possibly
in a few years we will refer to Figure 5 as the typical
radial BSS distribution in stellar aggregates.
In the meanwhile let us still   refer to it
as {\it peculiar} until its generality can be finally
addressed.

We are now exploiting the wealth of information in
the large dataset presented here by 
 performing an extensive
spectroscopic survey of the BSS population in 47 Tuc using the new
multi-fiber Spectrograph FLAMES at the ESO-Very Large Telescope.
These data should be highly informative. For instance, radial velocity
determinations for a significant number of EBSS would help to
clarify their origin: if the EBSS were ejected from the cluster
center their velocity dispersion should be low because they are
expected to be observed near the apocenter of an highly elliptical
orbit; if EBSS arise from primordial binaries the velocity
dispersion might be close to the Keplerian velocity (circular orbits
which avoid the cluster center might allow the binaries to remain in the
exterior) and show small variations due to low mass companions. The
same spectra could lead to many mass estimates: so far the only BSS
with a measured mass ($1.7 \pm 0.4\,\msun$) is in 47~Tuc and was
obtained with \hst\ \citep{shara-bssmass}.

\acknowledgements

We  warmly thank Paolo Montegriffo for assistance during
the astrometry procedure and Michela Mapelli and Andrea
Possenti for useful discussion. The financial  support of
the  Agenzia Spaziale Italiana (ASI)    and the Ministero
dell'Istruzione, dell'Universist\`a e della Ricerca  is
kindly acknowledged. This research has made use of  the
GSCII catalog which  has been produced by the Space
Telescope Science Institute and the Osservatorio
Astronomico di Torino and of  the ESO/ST-ECF Science
Archive facility which is a joint collaboration of  the
European Southern Observatory and the Space Telescope -
European Coordinating Facility. RTR is partially supported
by STScI grant GO-8709 and NASA LTSA grant NAG 5-6403.

\clearpage

\clearpage

\clearpage
\begin{figure} 
\plotone{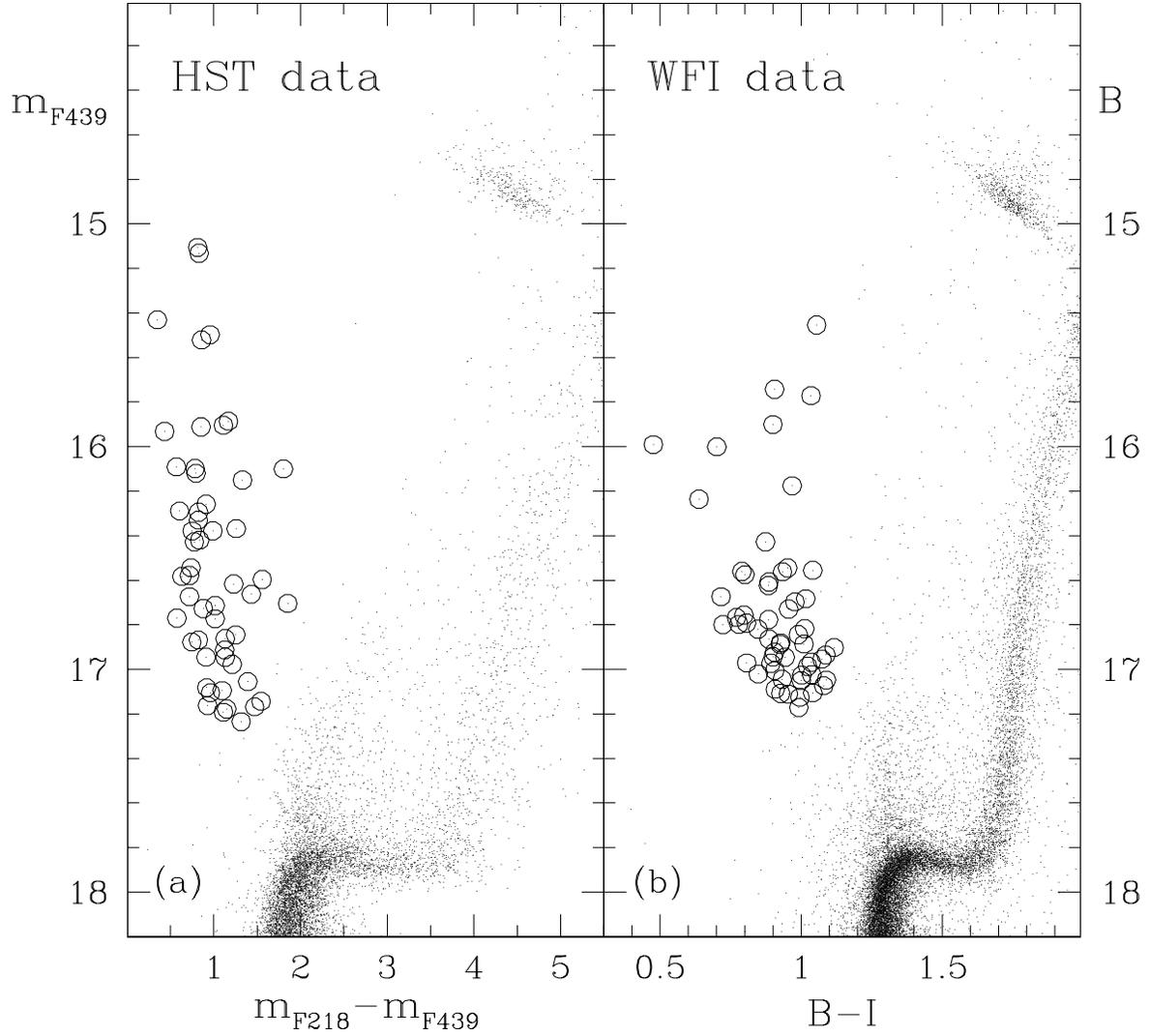} 
\caption[fig1.ps]{
Direct comparison between the HST ({\it Panel (a)}) and 
the ground-based WFI sample ({\it Panel (b)}). As in Figure
1 and 2 stars selected in the  BSS   population are marked
with empty circles. As can be seen the selected BSS 
samples are fully homogeneous.  
 \label{fig:CMD3}}
\end{figure}
  
\clearpage
 
\begin{figure} 
\plotone{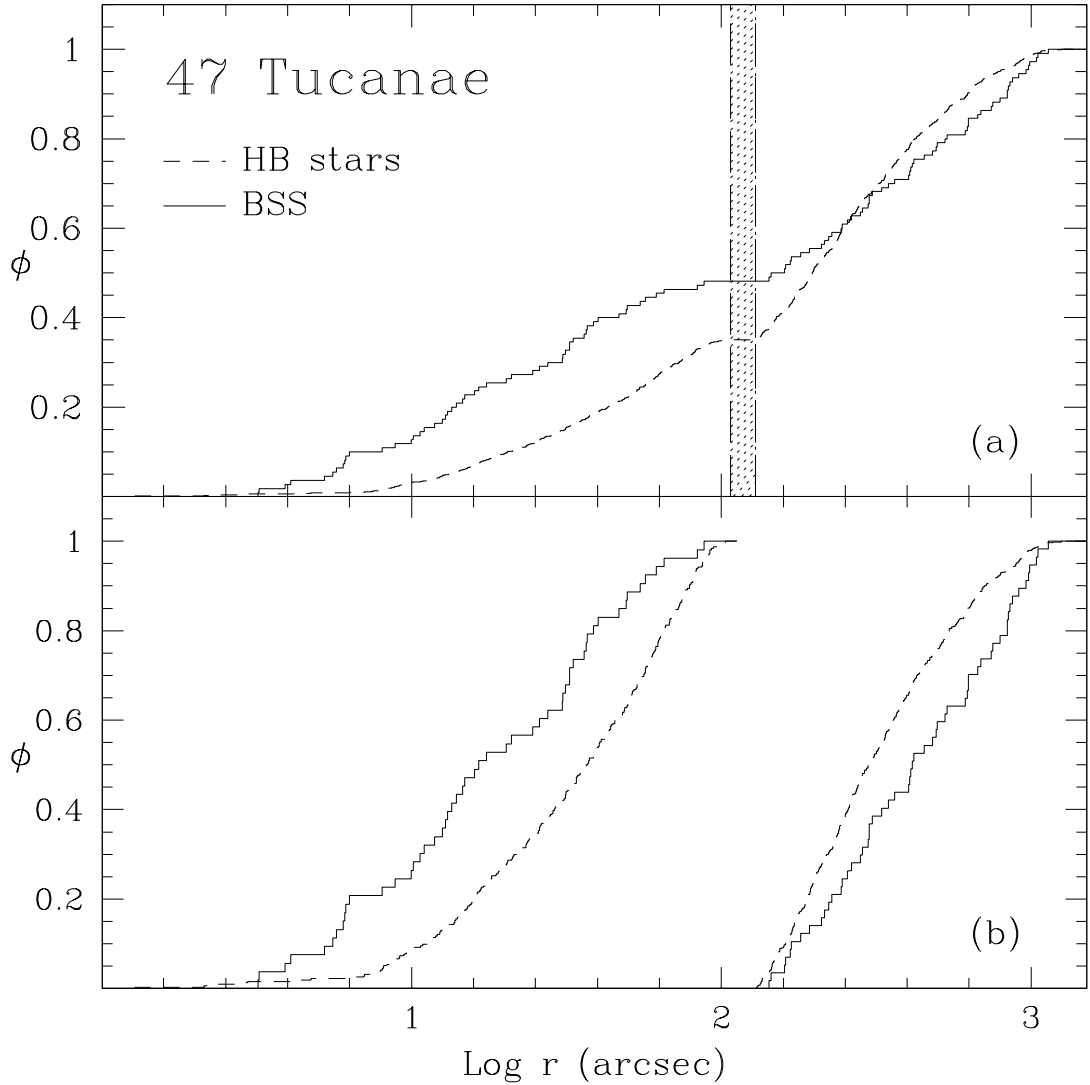}
\caption[fig2.ps]{{\it Panel (a):} 
Cumulative radial distributions for the {\em bright}  BSS
(solid line) with respect to the HB stars (dashed line) as
a   function of their projected distance   ($r$)  from the
cluster center for the global sample.  The vertical dotted
line distinguish the cluster regions observed   with HST
from the regions observed from the ground.  
{\it Panel (b):}  The same as {\it panel (a)} for two radial
subsamples.  
\label{fig:histogram}}
\end{figure}
   
\clearpage
\begin{figure} 
\plotone{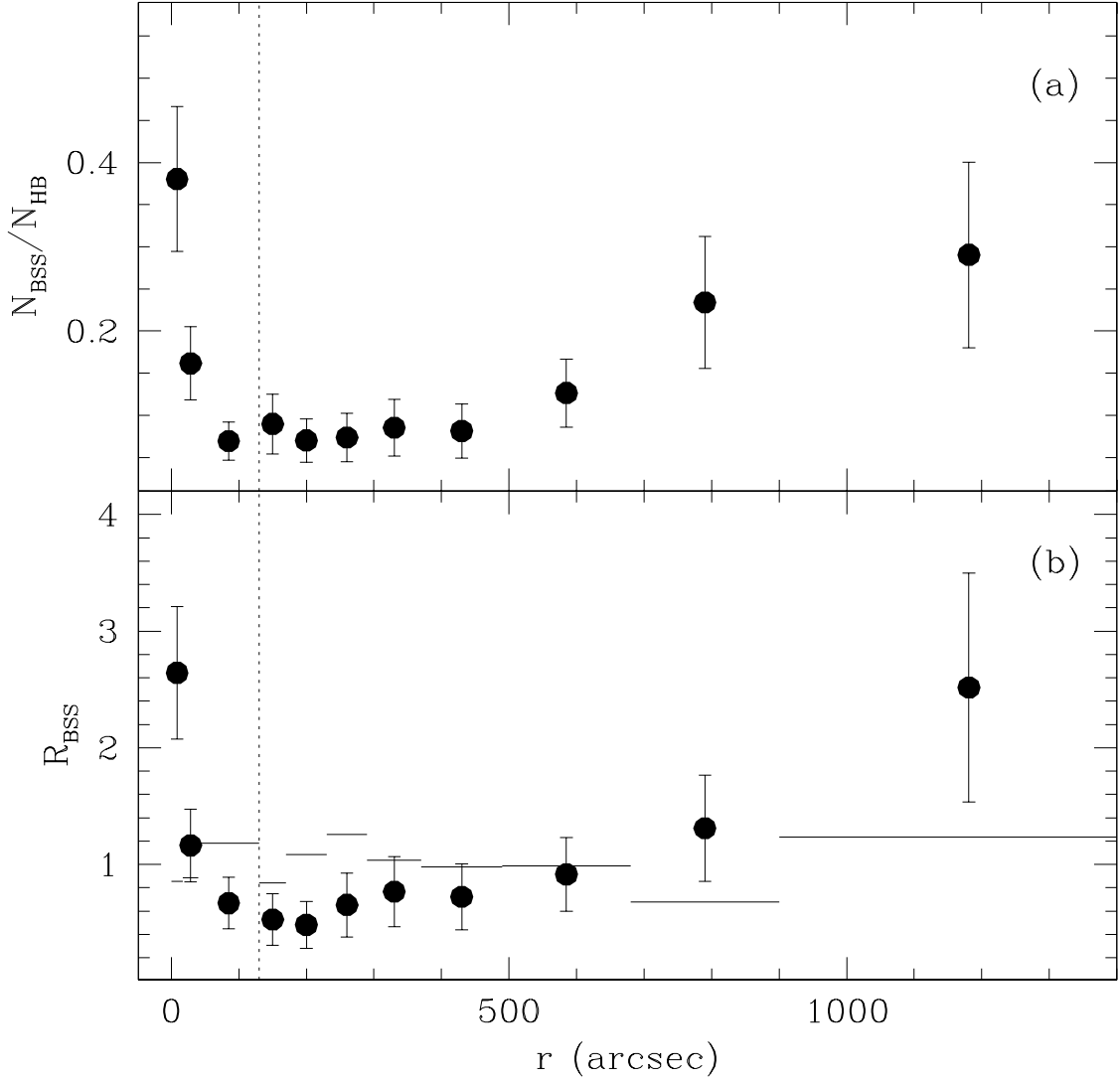}
\caption[fig3.ps]{ {\it Panel (a):} The relative frequency
of BSS with respect to HB stars is plotted as a function
of the distance from the cluster center for each of the 11
concentric annuli defined in the text.
 {\it Panel (b):} The specific frequency of BSS (see text) is
 plotted. The horizontal lines show the specific
 frequency for the HB reference population. 
 The vertical dotted line distinguish the cluster regions
observed   with HST from the regions observed from the
ground.  
\label{fig:NBSS}}
\end{figure} 
 
\clearpage 

\begin{figure} 
\plotone{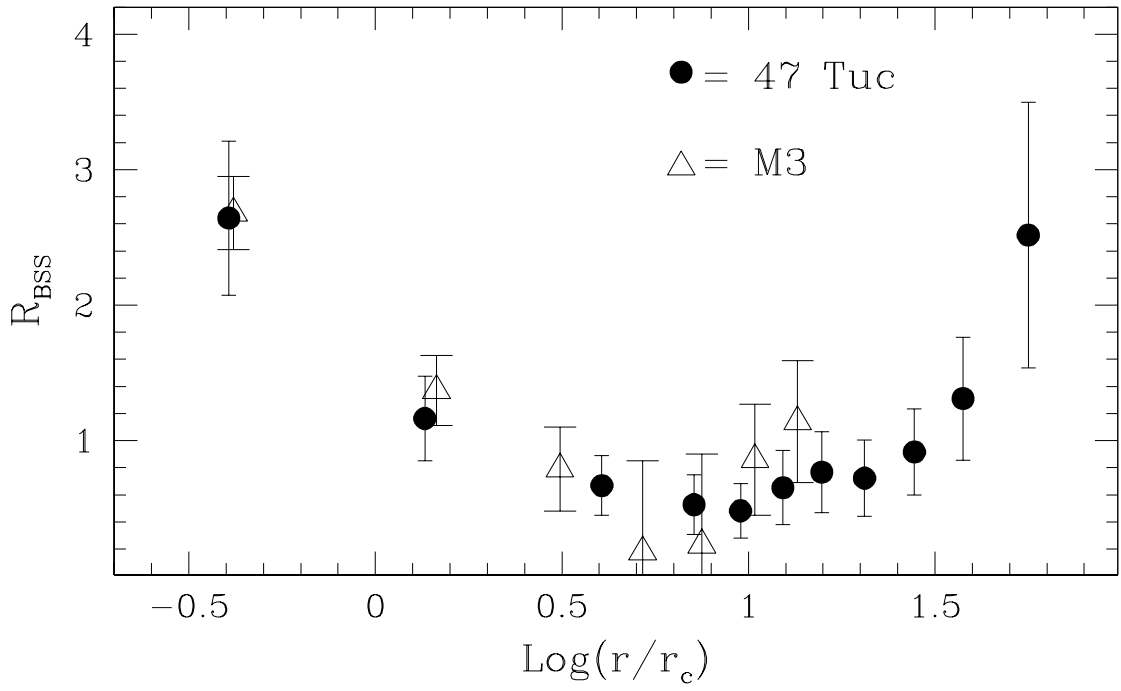} 
\caption[fig2.ps]{ The specific frequency of BSS in 47 Tuc
(filled circles) is compared with that found in M3 (empty
triangles). The radial coordinate is expressed in unit of
core radii $r_c$.  
\label{fig:cfM3}}
\end{figure}

\clearpage
  
\begin{deluxetable}{ccccccccc}
\scriptsize
\tablewidth{16cm}
\label{tab:uve}
\tablecaption{Numbers and relative frequencies for BSS and
HB stars.}
\tablehead{
\colhead{r} &
\colhead{$N_{\rm BSS}$} &
\colhead{$N_{\rm HB}$} &
\colhead{$N_{\rm BSS}/N_{\rm HB}$} &
\colhead{$L_{\rm annulus}/L_{\rm TOT}$} &
\colhead{ $R_{\rm BSS}$}& 
\colhead{  $\epsilon_{\rm BSS}$} &
\colhead{ $R_{\rm HB}$}&
\colhead{$\epsilon_{\rm HB}$}\\
}
\startdata
 0''-17''    & 27 & 71  & 0.38 &0.09 &  2.64  & 0.57  &0.85 &0.11 \\ 
17''-40''    & 16 & 99  & 0.16 &0.12 &  1.16  & 0.31  &0.88 &0.09 \\ 
40''-130''   & 10 & 144 & 0.07 &0.14 &  0.67  & 0.22  &1.18 &0.11 \\ 
130''-170''  & 6  & 78  & 0.09 &0.10 &  0.53  & 0.22  &0.84 &0.10 \\ 
170''-230''  & 6  & 110 & 0.07 &0.11 &  0.48  & 0.20  &1.09 &0.11 \\
230''-290''  & 6  & 94  & 0.07 &0.08 &  0.65  & 0.27  &1.26 &0.14 \\ 
290''-370''  & 7  & 77  & 0.09 &0.08 &  0.77  & 0.30  &1.04 &0.12 \\
370''-490''  & 7  & 77  & 0.08 &0.09 &  0.72  & 0.28  &0.98 &0.12 \\
490''-680''  & 9  & 79  & 0.13 &0.09 &  0.92  & 0.32  &0.99 &0.12 \\
680''-900''  & 9  & 38  & 0.23 &0.06 &  1.31  & 0.45  &0.68 &0.11 \\
900''-1462'' & 7  & 28  & 0.29 &0.03 &  2.52  & 0.98  &1.24 &0.24 \\
\enddata
\end{deluxetable}

\end{document}